\documentclass[twocolumn,final]{svjour3}
\usepackage[pdftex]{graphicx}  % Package for importing postscript images
\usepackage{color}
\usepackage[authoryear, round, numbers]{natbib}  % Package for Bibtex
\usepackage{amsmath}  % Package for typesetting formulae
\usepackage{hyperref}  % Package for hyperlinking
\usepackage[top=1.5in, bottom=1.5in, left=1in, right=1in]{geometry}
\usepackage{amsfonts}
\usepackage{verbatim}

\usepackage{booktabs} % Top and bottom rules for tables

\newcommand{\bm}[1]{\mbox{\boldmath $#1$}}

\title{Bayesian spectral density estimation using P-splines with quantile-based knot placement}
\author{Patricio Maturana-Russel \and Renate Meyer}

\institute{Patricio Maturana-Russel \at Department of Mathematical Sciences, Auckland
	University of Technology, Auckland, New Zealand \\ \email{p.maturana.russel@aut.ac.nz} 
\and Renate Meyer  \at Department of Statistics, University of Auckland, Auckland, New Zealand 
}

\newcommand{\pmr}{ \color{black}}

\begin{document}\sloppy  

\maketitle

\begin{abstract}

This article  proposes a Bayesian approach to estimating the spectral density of a stationary time series using a prior based on a mixture of P-spline distributions.  Our proposal is motivated by the B-spline Dirichlet process prior of \cite{Edwards2019} in combination  with Whittle's likelihood and aims at  reducing the high computational complexity of its posterior computations. The strength of the B-spline Dirichlet process prior over the Bernstein-Dirichlet process prior of \cite{Choudhuri:2004}  lies in its ability to estimate spectral densities with sharp peaks and abrupt changes due to the flexibility of B-splines with variable number and location of knots. Here, we suggest to use P-splines of \cite{Eilers:1996} that combine a B-spline basis with a discrete penalty on the basis coefficients.  In addition to equidistant knots, a novel strategy for a more expedient placement of  knots is proposed that makes use of the information provided by the periodogram about the steepness of the spectral power distribution.
We demonstrate in a simulation study and two real  case studies that this approach retains the flexibility of the B-splines, achieves similar ability to accurately estimate peaks due to the new data-driven knot allocation scheme but significantly reduces the computational costs.

\end{abstract}


\section{Introduction}
The power spectral density (psd), or simply spectral density, describes the distribution of the power or variance over the individual frequency components of a time series.  It embodies useful information for the study of stationary time series.  For a zero-mean weakly stationary time series, an absolutely summable autocovariance function, i.e.\ $\sum_{h=-\infty}^{\infty} |\gamma(h)| <\infty$, guarantees the existence of a continuous and bounded spectral density function given by
\begin{align*}
f(\lambda) = \dfrac{1}{2\pi} \sum_{h = -\infty}^{\infty}\gamma(h)\exp\left(-i h \lambda \right), 
\end{align*}
where $-\pi < \lambda \leq \pi$ is the angular frequency.

Methods to estimate this function range from parametric to nonparametric approaches.  The former are mainly based on fitting  autoregressive moving average (ARMA) models. However, the corresponding spectral density will be biased if the ARMA model does not adequately describe the dependence structure of the time series. This is a downside of all parametric approaches.
They are efficient if the model fits well but because they make restrictive assumptions about the data-generating mechanism, their inference is sensitive to model misspecifications and may  be severely biased. 
Bayesian nonparametric approaches, on the other hand, put a prior distribution on the set of all possible densities that could have generated the data not just a small parametric subset thereof. They thus avoid the problem of choosing between different parametric models  by an intrinsic data-driven determination of the model complexity.

As the periodogram fluctuates around the true spectral density, most  nonparametric methods for spectral density estimation hinge on  smoothing the periodogram
\begin{align*}
I_n(\lambda_l) = \dfrac{1}{2 \pi n} \left| \sum_{t=1}^{n} Y_t \exp \left( -i t \lambda_l\right)\right|^2,
\end{align*}
in one way or another,{\pmr where $ \lambda_l=2\pi l/n$ for $l=0,\ldots,\nu = \left\lfloor (n-1)/2 \right\rfloor$ are the Fourier frequencies} and $\bm{Y}=(Y_1,\ldots,Y_n)$ is a stationary time series of length $n$. The periodogram ordinates evaluated at the Fourier frequencies  are asymptotically independent and follow exponential distributions with mean{\pmr$f(\lambda_l)$}.  Thus, they are asymptotically  unbiased estimates of{\pmr$f(\lambda_l)$}  but not consistent \citep{Brockwell:1986}.

Most Bayesian approaches to psd estimation use the so-called Whittle likelihood, introduced by \cite{Whittle:1957} as an approximation to the likelihood of a Gaussian stationary time series.  Based on the  limiting independent exponential distribution of the periodogram values, the Whittle likelihood for a mean centred weakly stationary time series $\bm{Y}$ of length $n$ is defined as
\begin{align}
L(\bm{Y} | f) &= \prod_{l=1}^{\nu}\dfrac{1}{f(\lambda_{l})} e^{-I_n(\lambda_l)/f(\lambda_l)}  \nonumber \\
&\propto \exp \left\{ -\! \sum_{l=1}^{\nu} \Bigg( \log f(\lambda_l) + \dfrac{I_{n}(\lambda_l)}{f(\lambda_l)} \Bigg) \right\}.\label{eq:whittle_like}
\end{align}
%where $\lambda_l = 2\pi l / n$ are the Fourier frequencies, $f(\cdot)$ spectral density of $\bm{Y}$, and $\nu = \left\lfloor (n-1)/2 \right\rfloor$ is the %greatest integer value less than or equal to $(n-1)/2$.
 Advantages of this approximation are that it depends directly on the spectral density, unlike the true Gaussian likelihood function, and gives a good approximation for non-Gaussian time series under certain conditions as shown by \cite{ShaoXiaofeng2007ASTf}.

Bayesian nonparametric prior specifications for the psd include \cite{Cart:1997}
who used a smoothing prior on the log psd.  \cite{Rosen:2012} segmented a nonstationary time series into stationary segments.  For each stationary segment, they decomposed the log psd into a linear and nonlinear component and  put a linear smoothig spline prior on the nonlinear part.  \cite{Pensky:2007} propose Bayesian wavelet-smoothing of the log psd and recently, \cite{Cadonna2017} modelled the log psd  as a mixture of Gaussian distributions with frequency-dependent weights and mean functions.
\cite{Gangopadhyay:1999} modelled the log-spectral density function by  piecewise polynomials of low order between strategically placed knots on the support of the psd,  allowing the data to select the number and location of the knots.  A reversible jump Markov chain Monte Carlo \citep[RJMCMC;][]{Green:1995} algorithm was implemented to sample from the posterior distribution.  Similarly, frequentist approaches such as \cite{Rodriguez} and \cite{Wood2017} model the log spectral density by a linear combination of splines with variable knots and combine these with adaptive smoothing. 

Penalized splines have been used for a time-domain analysis of time series data e.g.\ to model the mean as a smooth function of time by \cite{Krivobokova}  or for fitting and  forecasting univariate  nonlinear time series by \cite{Wegener}, but not in the frequency domain for estimating the spectral density.
The P-spline prior on the spectral density function that we are going to propose is related to the 
Bernstein-Dirichlet process (BDP) prior  introduced by \cite{Choudhuri:2004} and combined with the Whittle likelihood.  The BDP prior  is a mixture of beta densities with weights induced by a Dirichlet process.  The number of components, which is a smoothing parameter, is given a discrete prior.  \cite{Edwards2019} extended the method by replacing the beta densities with B-spline densities. Similar to 
\cite{Gangopadhyay:1999}, the specification of the number and location of the B-spline densities is data-driven, however it avoids a RJMCMC algorithm by putting a second Dirichlet process prior  on the cdf that induces the inter-knot distances. \cite{Choudhuri:2004} already demonstrated in a comprehensive simulation study  that the Bayesian psd estimates based on the BDP prior outperforms in terms of the $L_1$-error the smoothed periodogram estimates based on Bartlett-Priestley quadratic kernel, the penalized MLE of \cite{Pawitan} and autoregressive spectral estimates in most scenarios but noted that the BDP-based estimates detect peaks correctly but often underestimate the magnitude of sharp peaks.
Unlike the beta densities  which have global support on the interval [0,1], B-spline densities have local support, thereby increasing their flexibility and allowing better modelling of spectral densities with sharp peaks and abrupt changes.  As demonstrated in \cite{Edwards2019}, the B-spline--Dirichlet process (BspDP) prior outperforms the BDP prior in estimating complex psds.  However, this flexibility comes at a high computational price.

%The Bernstein polynomial and the B-spline methods make use of Whittle's approximation to the Gaussian likelihood, known simply as Whittle likelihood \citep{Whittle:1957}.  

With the aim of reducing the computational cost of the BspDP prior approach,  we suggest to use the idea of P-splines of \cite{Eilers:1996} that combines a large but fixed number of  B-splines with a simple difference penalty that controls the degree of  smoothness of the spectral density. P-splines, i.e.\  equally-spaced B-splines,  have been successfully implemented for frequentist nonparametric regression but to the best of our knowledge not yet for Bayesian spectral density estimation. We suggest two novel Bayesian  approaches in this paper, both based on the Whittle likelihood and on a mixture of B-spline densities. Both approaches fix the  number of B-spline densities.  The first approach uses equally spaced knots with a P-spline prior which we define in Section III. We demonstrate that this approach yields significant savings in computational time without sacrificing accuracy over the BspDP prior approach for simple spectral structures.  The second approach has a particular  focus on spectral densities with sharp peaks. Instead of an equidistant spacing, the knots are spread based on the distribution of peaks in the periodogram. Thus the knot allocation is also data-driven as with the BspDP prior but the knots  remain fixed once allocated. This novel knot allocation scheme avoids the computational complexity of the BspDP prior.  We show that with this knot allocation scheme, the estimates of spectral densities with spikes and abrupt changes improves significantly without an increase in computing time.

%Motivated by the flexibility of the B-spline prior and its associated high computational cost, in this work, we implement and test a particular case of the B-spline prior.  For this, we fix the %number of B-spline densities and place a P-spline prior on the spectral density instead.  This approach allows to penalize automatically the inclusion of B-spline densities and save significantly %computational time without sacrificing accuracy.  Actually, as will show, it outperforms the B-spline prior in our tests.  In addition, we propose a knot location scheme based on the %periodogram, which aims to deal with psd shapes that possess  abrupt peaks, without increasing computational time.  We show in an example how this scheme improves significantly the %estimates.

The paper is organized as follows: In order to set notation and allow for a direct comparison, we start by
reviewing the B-spline densities and the BspDP prior in Section II. Section III specifies the P-spline prior for the psd, the novel knot allocations scheme and the Gibbs sampler used to sample from the posterior distribution based on the P-spline prior and Whittle likelihood.  In Section IV, we test our P-spline-based approaches to spectral density estimation and compare results to those based on the BspDP prior in a simulation study and the application to two real time series. The paper concludes with a discussion of the relative merits of the new approaches and avenues for future research.

\section{Notation and review of the B-spline--Dirichlet process prior}

By an application of the Weierstrass theorem it can be seen that a mixture of beta densities with only integer parameters can uniformly approximate any continuous density  on the interval $[0,1]$ \citep{Choudhuri:2004}.  Let $G$ be a cumulative distribution function (cdf), with continuous density $g$, then 
\begin{align}
\label{eq:Bernstein}
\widehat{g}(\omega) &= \sum_{k=1}^{K} G \left( \dfrac{k-1}{K} , \dfrac{k}{K} \right] \beta(\omega; k, K-k+1)\\
&= \sum_{k=1}^{K} w_k \beta(\omega; k, K-k+1) \nonumber
\end{align}	  
converges uniformly to $g(\omega)$, where $G(u,v] = G(v) - G(u)$, $\beta(\omega; a,b)$ is the beta density with parameters $a$ and $b$, and $w_k=G\left((k-1)/K, k/K \right]$ are the weights of the mixture.
The Bernstein polynomial prior, which is used to describe a nonparametric prior for probability densities in the unit interval, is based on this approximation \citep{Petrone:1999a,Petrone:199b}.  \cite{Choudhuri:2004} proposed it as a prior on the spectral density function.

{\pmr As shown by \cite{Perron:2001} for distributions on the interval [0,1], both mixtures of B-spline and beta distributions can approximate these arbitrarily well by increasing the number of mixture components, i.e.\ increasing the polynomial order for the beta distributions and the number of knots for suitable knot locations for the B-splines. But the rate of approximation is faster for B-splines than for beta distributions.}
%As shown by \cite{Perron:2001} for distributions on the interval [0,1], the mixture of B-spline distributions has better approximation properties than the mixture of beta distributions for a fixed number of components. In particular, the approximation error of the mixture of B-spline distributions can be made arbitrarily small by finding  suitable knot locations and increasing the number of knots {\pmr\citep{Perron:2001}}.
This work motivated \cite{Edwards2019} to propose the BspDP prior on the spectral density function which is based on the representation (\ref{eq:Bernstein}) but with
the beta densities replaced by B-spline densities as defined in the following.

\subsection*{B-splines and BspDP Prior}

A spline of {\pmr order} $r+1$ is a function defined piecewise by polynomials of degree $\leq r$, which meet at points called \textit{knots} where the function is continuous.  Any of these spline functions can be described by basis functions, known as \textit{B-splines}, in other words, all spline functions can be represented as a unique combination of B-splines with the same order and over the same partition.  Without loss of generality, the global domain of interest is assumed to be the unit interval $[0,1]$.

The B-spline basic functions of any order can be recursively defined as
\begin{align*}
B_{k,0}(\omega, \bm{\xi})=&	
\left\{
\begin{array}{ll}
1, & \omega \in [\xi_{k-1}, \xi_{k}] \\
0, & \mbox{otherwise} 
\end{array}
\right.\\
B_{k,r}(\omega, \bm{\xi}) = & \:v_{k,r}B_{k,r-1}(\omega; \bm{\xi}) + \\
& \: (1-v_{k+1,r})B_{k+1,r-1}(\omega; \bm{\xi}),	
\end{align*}
where	   
\begin{align*}	   
v_{k,r}&=	
\left\{
\begin{array}{ll}
\dfrac{\omega - \xi_{k-1}}{\xi_{k+r-1} - \xi_{k-1}}, & \xi_{k-1} \neq \xi_{k+r-1}\\
0, & \mbox{otherwise} 
\end{array}
\right.
\end{align*}
and 
\begin{align*}
\bm{\xi} = \{0 = \xi_0 = \xi_1 = \cdots = \xi_r \leq \xi_{r+1} \leq \cdots \leq \xi_{K}\\ = \xi_{K+1} = \cdots = \xi_{K+r} = 1 \}
\end{align*}
is the knot vector.  This vector is a non-decreasing sequence that contains $K+r+1$ knots, which can be divided in $2r$ external and ${\pmr K^*}=K-r+1$ internal knots.  The latter must be $\geq r$.
% external and internal knots as in Edwards et al (2018)

%This vector is a non-decreasing sequence that contains $K+r+1$ knots, which can be divided in $2(r+1)$ external and $K-r-1$ internal knots.  The latter must be $\geq r$.

The \textit{B-spline densities}, i.e.\ normalized B-spline functions, are defined by
\begin{align*}
b_{k,r}(\omega; \bm{\xi}) = \left(r+1\right) \dfrac{B_{k,r}(\omega;\bm{\xi})}{\xi_{k+r} - \xi_{k-1}}.
\end{align*}

The BspDP prior proposed by \cite{Edwards2019}
 involves two  Dirichlet processes, one  placed on $G$ inducing  the weights of the B-spline densities and the other on the knot spacings.  In practice, these are implemented using   the  stick-breaking representation \citep{Sethuraman:1994}, for details see \cite{Edwards2019}. 
% This method consists basically in a representation of the Dirichlet process as a sequential division of the unit interval infinite times according to a beta distribution. 
For computational reasons,  this infinite series representation of $G$ and $H$ is truncated at a large but finite positive integer, $L_G$ and $L_H$, respectively.  
These truncations control the quality of the approximations.  Longer series improve the approximation, but increase the number of calculations and consequently the computation time.

As shown by \cite{Edwards2019}, the approach based on the  BspDP prior outperforms the BDP prior in estimating spiky spectral densities.  In the case of smooth
spectral densities, there is no a significant difference in the performance of these two approaches.  However, in practice, the true psd function is unknown, which makes it necessary to use   methods such as the BspDP that will work well in all conceivable cases.

To implement the BDP prior, one only has to  calculate the beta densities once and they can be stored and re-utilized across MCMC steps.  On the other hand, the BspDP prior requires the calculation of the B-spline densities at each iteration, since these vary due to varying knot locations.  Moreover,  $2\!\times\!(L_G + L_H)$ calculations are needed in the truncated stick-breaking representation of the two Dirichlet processes  at each iteration %(see Eq. \eqref{eq:prior})
resulting in a significantly increased complexity of  the  BspDP algorithm.

This has prompted us to explore the performance of two new algorithms, both based on the P-spline prior algorithm. Both use a fixed number of B-spline densities and knot locations, avoiding the Dirichlet processes and the recalculation of the B-spline densities. Whereas the first approach uses  equidistant knots, the second algorithm  proposes a new fast knot allocation strategy. Both methods preserve the flexibility of the B-spline densities and their potential ability to estimate spectral densities with sharp peaks and abrupt changes.

%%%%%%%%%%%%%%%%%
%%% P-splines %%%
%%%%%%%%%%%%%%%%%

\section{P-spline prior}

As the spectral density function $f(\cdot)$ is defined on the interval $[0,\pi]$, it is reparametrized and defined as
\begin{align*}
f(\pi \omega) =\tau \times s_r(\omega;\bm{w}, \bm{\xi}), \quad \omega \in [0,1], %\tau \times s_r(\omega;K,G,H), \quad \omega \in [0,1],
\end{align*}
where $\tau = \int_{0}^{1}f(\pi \omega)\text{d}\omega$ is the normalizing constant and
\begin{align}
s_r(\omega; \bm{w}, \bm{\xi}) = {\pmr s_r(\omega)} = \sum_{k=1}^{K} w_k b_{k,r}(\omega;\bm{\xi}),
\label{eq:ps_prior}	
\end{align}
with  {\em fixed} number $K$ of B-spline densities  $b_{k,r}(\cdot)$ of fixed degree $r$, weight vector $\bm{w}=(w_1,\cdots, w_K)$, and knot sequence $\bm{\xi}$.

%Fixing the number of B-spline densities $K$ in Eq. \eqref{eq:bs_prior}, the prior for the unit interval $[0,1]$ is defined as
%\begin{align*}
%s_r(\omega;\bm{w}, \bm{\xi}) = \sum_{k=1}^{K} w_{k} b_{k,r}(\omega;\bm{\xi}),	
%\end{align*}
%where $\bm{w} = (w_1,\dots,w_K)$.

$\bm{\xi}$ contains fixed equidistant internal knots on $[0,1]$ in our basic P-spline prior.  A more judicious scheme for the placement  of knots that improves the fit for peaked spectral densities is described in  section ``Quantile-based knot placement".
%Therefore, the P-spline prior on the spectral density function is given by
%\begin{align*}
%f(\pi \omega) = \tau \times s_r(\omega;\bm{w}, \bm{\xi}), \quad \omega \in [0,1],
%\end{align*}
%where $\tau = \int_{0}^{1}f(\tau \omega)\text{d}\omega$ is the normalizing constant.

Smoothing splines in the frequentist context are based on approximating a given function by a linear combination of B-splines and
putting a penalty on the integrated squared second derivative  as a measure of roughness. P-splines avoid derivatives by expressing the penalty as the sum of squares of differences of the B-spline coefficients.
%The roughness penalty in a frequentist context is carried out by penalizing the likelihood function.  
In a Bayesian context, this penalty can be transformed into a prior distribution for the $d^{\text{th}}$ order difference of successive coefficients, see e.g. \cite{Lang:2004}.
%, known also as B-spline parameters.  

In our approach, the B-spline coefficients $w_k$ are weights, so positive and sum to one. 
We therefore reparametrize to the vector $\bm{v}$ with
\begin{align*}
v_k = \log \left( \dfrac{w_k}{1-\sum_{k=1}^{K-1}w_k} \right).
\end{align*}
After some calculations, it can be shown that the weights are given by
\begin{align*}
w_k = \dfrac{e^{v_k}}{1+ \sum^{K-1}_{k=1}e^{v_k}}.
\end{align*}
The last weight is defined as $w_K = 1 - \sum_{k=1}^{K-1}w_k$, thus they all sum to 1.
Then, an indirect prior is placed on the weights via
\begin{align*}
\bm{v}|\phi&\sim\text{N}_{K-1}(\bm{0}, (\phi \bm{P})^{-1})\\
\phi|\delta &\sim \text{Gamma}(\alpha_{\phi}, \delta \beta_{\phi})\\
\delta &\sim \text{Gamma}(\alpha_{\delta}, \beta_{\delta})
\end{align*}
where $\bm{v} = (v_1,\dots,v_{K-1})^\top$ is a $K-1$ dimensional parameter vector, $\phi$ is the smoothing or penalty parameter, $\textbf{P} = \textbf{D}^\top \textbf{D} +\epsilon \textbf{I}_{K-1}$ is the penalty matrix (which is a full matrix rank matrix for any small quantity $\epsilon$, for instance, $10^{-6}$) with $\textbf{D}$ the $d^{\text{th}}$ order difference matrix ($d=1,2,\ldots, K-2)$, $\alpha_{\phi}$ and $\alpha_{\delta}$ are shape parameters, $\delta \beta_{\phi}$ and $\beta_{\delta}$ are rate parameters, and $\text{Gamma}(a,b)$ denotes a gamma distribution with mean $a/b$ and variance $a/b^2$.
We used first and second order difference penalties in the simulation study.
The $1^{\text{st}}$ order difference matrix \textbf{D} $\in \mathbb{R}^{(K-2)\times (K-1)}$  is defined as
\begin{equation}\label{eq:D1}
%\begin{align*}
\begin{bmatrix}
-1 & 1   & 0  & 0 & \cdots & & & 0 \\
0  &  -1 & 1  & 0 & \cdots & & & 0 \\
\vdots  &    & \ddots & \ddots   & &&& \vdots \\   
0 &    &    &   & \cdots & 0 &  -1 & 1 \\
\end{bmatrix}
%\end{align*}
\end{equation}
and the $2^{\text{nd}}$ order difference matrix $\in \mathbb{R}^{(K-3)\times (K-1)} $ as
\begin{equation}\label{eq:D2}
%\begin{align*}
\begin{bmatrix}
1 & -2 &  1 & 0 & 0 & \cdots & & & & 0 \\
0 & 1  & -2 & 1 & 0 & \cdots & & & & 0 \\
\vdots & & \ddots & \ddots & \ddots &&&&& \vdots \\
0 & &&&& \cdots & 0 & 1 & -2 & 1 
\end{bmatrix}.
%\end{align*}
\end{equation}
%and for the $3^{\text{rd}}$ order is
%\begin{align*}
%\setcounter{MaxMatrixCols}{12} % to include more than 10 columns in bmatrix enviroment
%\begin{bmatrix}
%1 & -3 &  3 & -1 &  0 & 0 & \cdots & & & & & 0 \\
%0 &  1 & -3 &  3 & -1 & 0 & \cdots & & & & & 0 \\
%\vdots &    & \ddots &  \ddots & \ddots & \ddots & &  & & & & \vdots \\
%0 &   &  &   &  &  & \cdots & 0 & 1 & -3 & 3 & -1  
%\end{bmatrix}
%\end{align*}
%$\in \mathbb{R}^{(K-4)\times (K-1)}$.  
The rationale for using discrete second order roughness penalties as an approximation to the usual continuous roughness measure based on the integral over the second derivatives is given
in \cite{Eilers:1996}.

The parameter vector for the P-spline prior is $\bm{\theta} = (\bm{v}^\top, \phi, \delta, \tau)^\top$.
We use a robust specification for the  prior distribution of the penalty parameters as suggested by \cite{Jullion:2007} by choosing small values for $\alpha_{\delta}$ and $\beta_{\delta}$, for instance, $10^{-4}$. The  choice of $\alpha_{\phi}$ and $\beta_{\phi}$ do not affect the spectral density estimate. Here we set $\alpha_{\phi}$ and $\beta_{\phi}$ equal to 1 as used by \cite{Bremhorst:2016}. The scale parameter $\tau$ is given by an inverse gamma prior IG($\alpha_{\tau},\beta_{\tau}$), {\pmr which can be considered as a noninformative prior for $\alpha_{\tau} = \beta_{\tau} = 0.001$ \citep{Edwards2019}.}  These are also the prior specifications used in the simulations and examples in Section \ref{sec:application}.

\subsection*{Posterior Computation}

The  joint posterior distribution of the parameter vector $\bm{\theta} = (\bm{v}^\top, \phi, \delta, \tau)^\top$ is given by
\begin{align*}
p(\bm{\theta}|\bm{Y}) & \propto L(\bm{Y}|f) \times p(\bm{\theta})  \\
&= L(\bm{Y}|f) \times p(\bm{v}|\phi, \delta) \times p(\phi|\delta) \times p(\delta) \times p(\tau)
\end{align*}
where $ L(\bm{Y}|f) $ denotes the Whittle likelihood defined in \eqref{eq:whittle_like}.
This posterior is proper since all the prior distributions are proper.  
We use the Gibbs sampler to sample from the joint posterior distribution.

Sampling from the full conditional posterior distributions of $\phi$, $\delta$, and $\tau$ can be performed directly. These are given by
%The conditional posterior of $\phi$, $\delta$, and $\tau$ belong to known families of distributions, that is
\begin{align*}
\phi|\bm{Y},\delta,\bm{v} &\sim \text{Gamma}\left(\tfrac{K-1}{2} + \alpha_{\phi},\: \tfrac{1}{2} \bm{v}^\top \textbf{P} \bm{v} + \delta \beta_{\phi} \right), \\
\delta|\bm{Y},\phi  &\sim \text{Gamma}\left(\alpha_{\phi} + \alpha_{\delta}, \beta_{\phi} \phi + \beta_{\delta}\right),\:\: \text{and}\\
\tau|\bm{Y},\delta, \phi,\bm{v}   &\sim \text{IG}\!\left(\alpha_{\tau}+ \nu,\:  {\textstyle\sum\limits_{l=1}^{\nu}} \tfrac{I_n(\lambda_l)}{{\pmr s_r(\lambda_l/\pi)}} + \beta_{\tau} \right)\!.
\end{align*}
%The posterior distribution of $\tau$ is the same one obtained under the B-spline prior approach.

However, to sample from the full conditional distribution of $\bm{v}$, we use the Metropolis algorithm.
Since the weights in general are larger in those areas of the frequency domain where the psd has peaks, we specify their starting value proportionally to the periodogram. Applying the corresponding transformation we obtain the starting value for the vector $\bm{v}$.  This strategy speeds up the MCMC process by shortening the burn-in period.

In order to improve the mixing of the chains, we apply the Metropolis algorithm on a reparametrized posterior distribution \citep{Lambert:2007}.  For this, we need a pilot posterior sample for $\bm{v}$ from which we calculate its mean vector $\bm{\overline{v}}$ and covariance matrix $\textbf{S}$.  Then we define the following re-parametrization
\begin{align*}
\bm{v} = \textbf{S}^{1/2} \bm{\beta} + \overline{\bm{v}},
\end{align*}
where $\bm{\beta}=(\beta_1, \dots, \beta_{K-1})^\top$.

We can update $\bm{v}$ by modifying $\bm{\beta}$ according to a proposal distribution.  In this work, we propose a univariate proposal value $\beta_{k}^{*}$ to update $\bm{v}$ given by
\begin{align*}
\beta_{k}^{*} &= \beta_{k} + \sigma z,\\
\bm{v}^{*} &= \textbf{S}^{1/2}\bm{\beta}^{*} + \overline{\bm{v}}, 
\end{align*}
where $\bm{\beta}^{*}= (\beta_1, \dots, \beta_{k}^{*} ,\dots, \beta_{K-1})^\top$, $z \sim \text{Normal}(0,1)$, and $\sigma$ controls the length of the proposals.  We vary $\sigma$ across iterations in order to get an acceptance rate between 0.3 and 0.5.  Note that even though the proposal is univariate on $\bm{\beta}$, it is multivariate on $\bm{v}^*$.
Alternatively to the Metropolis step used for \bm{v}, a data augmentation step based on P\'olya-Gamma auxiliary variables can be used which would allow for a full Gibbs sampler \citep{Polson:2013}. 

The $m^{\text{th}}$ cycle of the Gibbs sampler is given by
\begin{itemize}
	\item Draw $\bm{v}^{m}$ from $p(\bm{v}|\bm{\beta}^{m-1},\phi^{m-1},\delta^{m-1},\tau^{m-1})$ using $K-1$ univariate Metropolis steps for $\beta_k$, with $k=1,\dots,K-1$, according to the reparametrization described above;
	\item Draw $\phi^{m}$ from 
	\begin{align*}
	\text{Gamma}\!\left(\tfrac{K-1}{2}\!+\!\alpha_{\phi},\: \tfrac{1}{2} \bm{v}^{m^\top} \textbf{P} \bm{v}^{m}\!+\!\delta^{m-1} \beta_{\phi} \right);
	\end{align*}
	
	\item Draw $\delta^{m}$ from $\text{Gamma}\left(\alpha_{\phi} + \alpha_{\delta}, \beta_{\phi} \phi^{m} + \beta_{\delta}\right);$ 
	\item Draw $\tau^{m}$ from 
	\begin{align*}
	\text{IG}\!\left(\alpha_{\tau}\!+\!\nu, {\textstyle\sum\limits_{l=1}^{\nu}} \tfrac{I_n(\lambda_l)}{{\pmr s_r(\lambda_l/\pi)}} + \beta_{\tau} \right) .
	\end{align*}
\end{itemize}

The influence of the penalty is determined by $\phi$.  The larger this value is, the smoother is the resulting estimate. %, whereas the lower the value, the rougher the results.
It is interesting to note the role of the penalty matrix  \textbf{P} in the rate parameter on the full conditional posterior distribution of $\phi$.  When the quadratic form $\bm{v}^\top \textbf{P} \bm{v}$ tends to infinity, this distribution becomes concentrated at $0^{+}$.  This limiting behaviour yields rougher results.  Note that in this case the prior rate $\delta\beta_{\phi}$ becomes irrelevant in the full conditional posterior density.  On the other hand, when $\bm{v}^\top \textbf{P} \bm{v}$ tends to $0^{+}$, the distribution favours large $\phi$ values, yielding smoother estimates.  
The magnitude of this quadratic form is controlled by the penalty order and the number of B-spline densities.  In general, large order penalties produce lower values for this quantity, causing smoother results.  In this way the penalty matrix penalizes the inclusion of B-spline densities.

\cite{Ruppert2002} explained that the degree of smoothness implied by a certain choice of prior parameters also depends on the scale of the responses $\bm{Y}$.  To avoid the problem, the author suggested to standardize $\bm{Y}$ before the sampling process and apply the inverse operation afterwards.  This also allows to avoid numerical problems in the MCMC process.  We follow his proposal.
%page 187   

\subsection*{Choosing the number of B-splines}

The number of B-spline densities plays a critical role in the model fit and may result in under- or oversmoothing.
%Too few could cause an underfitting whereas too many an overfitting.  
Even though the penalty parameter controls the smoothness of the fit, a large number of B-splines yields rougher and  a low number  smoother spectral density estimates.  There is a smallest sufficient number which results in a psd estimate that fits the features of the data whereas choosing a larger number will have an insignificant effect on the fit.   Unexpectedly, \cite{Ruppert2002} found some cases in which too many B-splines degrades the fitting in terms of mean square error.  Unfortunately, there is no  consensus in how to find the optimal value of the number of B-splines.  

\cite{Eilers2015} consider the use of $K=100$ B-splines a wise choice, unless computational constraints are evident.  Some algorithms that explore different number of knots in a trial sequence and find an optimal value with respect to a certain goodness-of-fit criterion have been proposed. \cite{Ruppert2002} suggested a full-search and myopic algorithms which are based on the generalized cross-validation statistic.  However, their results are not conclusive and this criterion only applies if the penalty parameter is treated as a tuning constant~\citep{Kauermann2011}.  \cite{Likhachev2017} also considered a knot sequence and proposed the selection of the number of knots based on  Akaike's, corrected Akaike's and
the Bayesian Information Criterion.

\cite{Ruppert2002} discussed a heuristic rule of thumb $\min\left\{n/4, 40\right\}$, which is simpler and, in our experience, seems to work very well in general.  This criterion allocates a knot between $\max\left\{ 4, \lfloor n/40 \rfloor \right\}$ observations, where $\lfloor \cdot \rfloor$ stands for the floor function.  We employ this approach in the application Section for selecting the number of B-spline densities.

% comment: in psd estimation the knots are alocated in the frequency domain. The knots are not allocated among the observations.  Maybe it should be commented.

%\subsection*{Knot allocation strategy}
\subsection*{Quantile-based knot placement}

The knots are often equally spaced in P-spline methods.  This scheme works quite well  for smooth psds without any abrupt changes.  However, for spectral densities with spikes, the algorithm will require a large number of knots to adequately estimate the peaks.  The B-spline prior method is able to handle abrupt changes because the knot location is variable and  driven by the data.  We propose a new selective method that allocates the knots according to quantiles of a periodogram-related distribution function but once allocated, the knots stay fixed.  The idea is to concentrate more effort in those regions that have potential peaks as detected by the periodogram.  It is similar in spirit to the knot placement of O-splines \citep{WandOrmerod} in the regression context that make use of the empirical quantiles of the covariates. 

Our knot location proposal works as follow.  We calculate the periodogram and apply a square root transformation in order to have more regular magnitudes and eliminate a potential trend.  Second, we standardize these values, apply the absolute value function,{\pmr and normalise them}.  
We treat this transformed periodogram as a{\pmr probability mass} function (pmf) and interpolate its distribution function by the continuous cumulative distribution function $F$.
Finally, the knots are allocated according to the quantiles of $F$.
 Thus, in those areas in which the periodogram has sharp and abrupt changes, our procedure will assign more knots in proportion to their magnitudes. Hereafter, we will refer to this scheme as {\em  quantile-spaced (Q-spaced) knots}.%\textit{periodogram-spaced knots} or short \textit{P-spaced knots}.
%REMARK: HOW ABOUT "PERIODOGRAM-SPACED KNOTS"	OR 'P-SPACED KNOTS'  INSTEAD OF DISTRIBUTED KNOTS?

%The procedure can be summarized as follows:
%\begin{enumerate}
%	\item Let $\bm{x} = \left (\sqrt{I_n(\lambda_1)}, \dots, \sqrt{I_n(\lambda_{\nu})} \right)$ be the square root of the periodogram values for each angular frequency.
%	\item Define $z_i = \left|\dfrac{x_i - \bar{x}}{S_x} \right|$ where $\bar{x}$ denotes the average and $S_x$ the standard deviation of ${\bm x}$ and calculate its distribution function $F_z = P(z \leq z_i)$.
%	\item Let $\bm{\xi}^* = \left(\xi_1^*,\dots,\xi_{K^*}^* \right)$ be a vector of $K^*$ quantiles, equally spaced in $[0,1]$.
%	\item The Q-spaced knot vector is given by $\bm{\xi} = \left(F_{z}^{-1}(\xi^*_1),\dots, F_{z}^{-1}(\xi^*_{K^*}) \right)$.	
%\end{enumerate}

{\pmr
The procedure can be summarized as follows:	%for the frequencies in the unit interval $\omega_l = 2 l/n  \in[0,1]$, which can be easily translated to the frequency domain, can be summarized as follows:	
\begin{enumerate}
	\item Let $\bm{x} = \left (\sqrt{I_n(\lambda_1)}, \dots, \sqrt{I_n(\lambda_{\nu})} \right)$ be the square root of the periodogram values.
	\item Define $y_l = \left|\dfrac{x_l - \bar{x}}{S_x} \right|$, for $l=1,\ldots,\nu$, where $\bar{x}$ denotes the average and $S_x$ the standard deviation of ${\bm x}$.
	\item Define $z_l = \frac{y_l}{\sum_{l=1}^{\nu} y_l}$, $Z_l= \sum_{j\leq l} z_j$, and let $F$ denote the continuous cdf that interpolates the $Z_l$.
	\item Let $\bm{q} =\left( q_1, \dots, q_{K^*} \right)$ be a vector of $K^*$ equally spaced points  in  $[0,1]$, that is $q_j = \frac{j-1}{K^*-1}$, for $j=1,\dots,K^*$.
	\item The Q-spaced internal knot vector is given by the corresponding quantiles $\bm{\xi}^* = \left(F^{-1}(q_1),\dots, F^{-1}(q_{K^*}) \right)$. %where the inverse function is obtained by interpolation.	
\end{enumerate}

Note that this knot allocation algorithm is applied only once at the beginning of the MCMC process.  Therefore, its computational cost is negligible.

}

For non-equidistant knots, it does not make sense any longer to base the definition of the covariance matrix of the Gaussian prior  on the  difference matrices as defined in (\ref{eq:D1}) and (\ref{eq:D2}) because this is an adequate approximation of the derivative-based roughness measures -- defined as the integral of the first and second order derivatives of the P-splines -- 
for equally spaced knots only. The definition of the difference matrices could be adjusted using divided differences, but here we use the derivative-based penalties as described by \cite{WoodSimon2017Pwdb}. We make use of the implementation of the derivative-based penalty matrix  by the function {\tt bsplinepen} in the R-package {\tt fda} \citep{fda}. This penalty matrix contains the inner products of the respective first and second order derivatives of the basis functions. In addition, we normalise this matrix by dividing it by the maximum absolute column sum.

The quantile based knot allocation can be regarded as an empirical Bayes prior since the data are used to specify the prior distribution.  However, only the locations of the B-splines are determined by the data, not the parameters of the prior distribution which in this case are the weights $w_k$ associated with each B-spline $b_{k,r}$.  The main criticism of an empirical Bayes approach is that the data are being used twice, once to specify the prior and then again to update the prior to the posterior. Here, the data are used only once to specify the knot locations but not again to update these through the likelihood.

%REMARK: CAN YOU PLEASE INSERT EXPLICIT FORMULAE.

\section{Application}
\label{sec:application}

We first assess the properties of our P-spline spectral density estimates in a simulation study and compare them to estimates based on the BspDP prior. We furthermore apply all approaches to the
analysis of the classical sunspot dataset  and the {\it S.\ Carinae} variable star light intensities, previously analysed by \cite{Cart:1997,Huerta:1999,Kirch:2018}.  As the psd of the  {\it S.\ Carinae} time series contains sharp peaks, its analysis allows to assess the impact of the periodogram-spaced knots in contrast to the equidistant knots.   For all the analyses, we use cubic B-splines ($r=3$).

\subsection{Simulation study}

\begin{figure}[]
	\centering
	\includegraphics[scale=0.4,clip=true,angle=0]{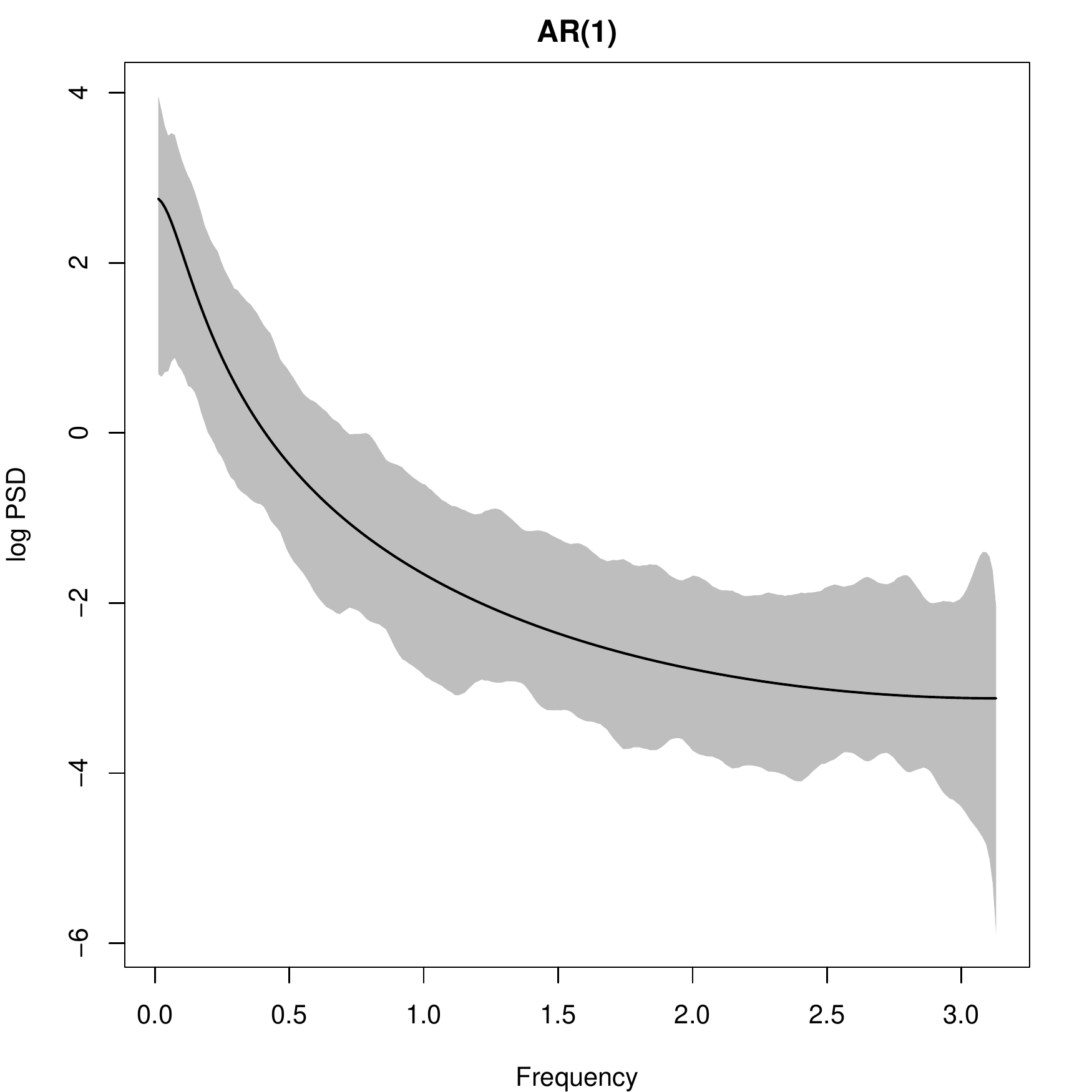}
	\caption{Estimated log-spectral density for an AR(1) time series via the P-spline prior algorithm with a first order penalty.  The solid line stands for the true psd, whereas grey area is the uniform 90\% credible band.}
	\label{fig:ar1}
\end{figure}

\begin{figure}[]
	\centering
	\includegraphics[scale=0.4,clip=true,angle=0]{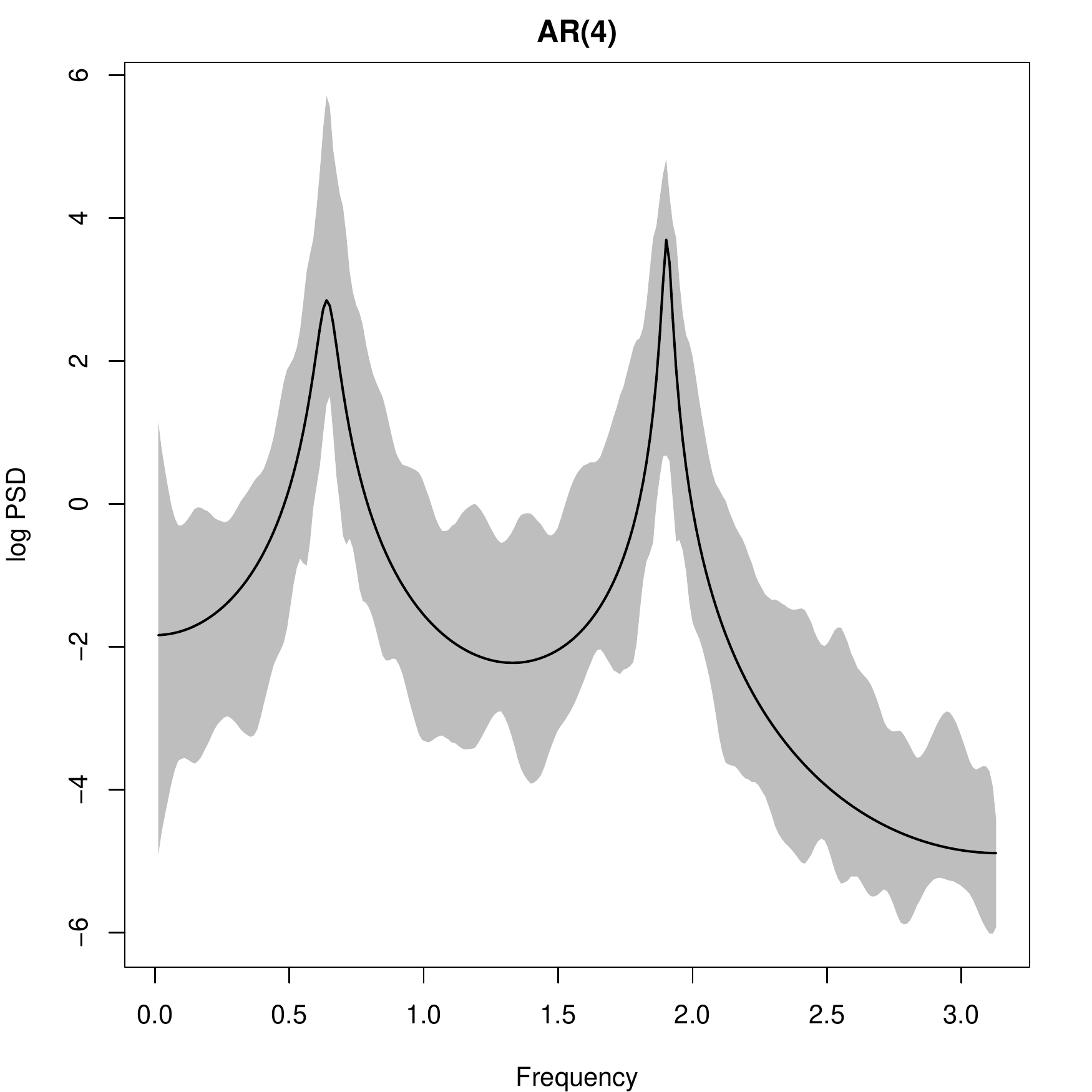}
	\caption{Estimated log-spectral density for an AR(4) time series via the P-spline prior algorithm with a first order penalty.  The solid line stands for the true psd, whereas grey area is the uniform 90\% credible band.}
	\label{fig:ar4}
\end{figure}

The setup of our simulation study mirrors that  in \cite{Edwards2019} who  compared spectral density estimates based on the BspDP and the BDP priors. 
% We perform the same analysis, varying only the number of iterations used in the analysis.  
We generated  300  autoregressive time series of order 1 and 4 with unit variance Gaussian innovations of length $n = \{128, 256, 512\}$.
For the AR(1) model, the  first  order autocorrelation $\rho_1 = 0.9$ was chosen, which has a relatively simple spectral density (see Figure~\ref{fig:ar1}).  The AR(4) time series with $\rho_1~=~0.9$, $\rho_2~=~-~0.9$, $\rho_3~=~0.9$ and $\rho_4~=~-~0.9$ have a spectral density with two large peaks (see Figure~\ref{fig:ar4}).  

We estimated the spectral density functions using the BspDP and P-spline priors.  The BspDP-based  analysis was performed via the \texttt{gibbs\_bspline} function in the \textsf{R} package \texttt{bsplinePsd{\pmr 0.6.0}} \citep{Edwards:bsplinePsd:2018}.  The algorithm was executed for 100,000 iterations with a burn-in period of 25,000 and thinning factor of 10, resulting in 7,500 samples used for posterior inference.  Pilot analyses showed that these specifications are suitable in these examples in terms of convergence (results not shown).  

The P-spline analysis was performed using the \texttt{gibbs\_pspline} function implemented in the \textsf{R} package \texttt{psplinePsd} \citep{psplinepackage}.  A total of 100,000 MCMC samples were generated
using  a pilot run of 20,000 iterations with a burn-in period of 5,000 and thinning factor of 10 to calibrate the proposals.  The final sample consisted of 80,000 samples with a burn-in period of 5,000 and thinning factor of 10, resulting in 7,500 samples used for posterior inferences.  
%Therefore, the number of iterations equate the B-spline analysis.  
This procedure was replicated for both equidistant and Q-spaced knots considering first and second order penalties.

The theoretical spectral density for an AR($p$) model is given by
\begin{align*}
f(\lambda) = \dfrac{\sigma^2}{2 \pi} \dfrac{1}{|1-\sum_{j=1}^{p}\rho_j \exp(-i \lambda)|^2},
\end{align*}
where $\sigma^2$ is the variance of the white noise innovations and ($\rho_1, \dots, \rho_p$) are the model parameters.  In order to compare the psd estimates  to the true function, we use the integrated absolute error (IAE) or $L_1$-error which is defined as
\begin{align*}
\text{IAE} = || \widehat{f} - f ||_1 = \int_{0}^{\pi} |\widehat{f}(\textcolor{red}{\lambda})-f(\textcolor{red}{\lambda})|\text{d}\textcolor{red}{\lambda},
\end{align*}
where $\widehat{f}(\cdot)$ is the pointwise posterior median of the spectral density $f(\cdot)$.   The IAE is computed for each of the 300 replications and its median is calculated.  
The results are displayed in Table~\ref{table:sim_IAE}.

For the AR(1) time series, the P-spline prior with equidistant knots and penalty order $d=1$ yields better psd estimates than the BspDP prior.
A larger penalty order does not necessarily result in more accurate psd estimates. In this case of a relatively smooth spectral density, the Q-spaced knots do not yield an improvement over the equidistant knots and the BspDP prior.

% yields more accurate results than the B-spline prior for $n=128$.  Slightly better results are obtained for $n=256$ with equidistant knots.  However, the distributed knot scheme yields %slightly higher values.  For $n=512$ and both knot schemes, only the first order penalty P-splines yield better results, whereas the second and third order penalty P-splines yield marginally %higher median IAE values.  In general, the equidistant knot scheme yield lower IAE values than the distributed knots.

For the AR(4) time series, the fit based on the P-spline prior with Q-spaced knots is better than the fit based on the B-spline Dirichlet process prior and the one based on the P-spline prior with equidistant knots.
% categorically outperforms the BspDP algorithm.   It also outperforms the equidistant knot scheme. 
 In general, the BspDP algorithm achieves higher accuracy than the P-spline method with equidistant knots.  As shown by \cite{Edwards2019}, the posterior distribution based on the BspDP prior yields better results than posterior based on the BDP in terms of median IAE in this example.  Consequently, our P-spline algorithm with Q-spaced knots outperforms the BDP method as well.

%The latter yields more accurate results than the BspDP prior for $n=\{128, 256\}$.  For $n=512$, the third order penalty P-splines has a slightly higher median IAE, but overall, this approach achieves higher accuracy.  
%As shown by \cite{Edwards2018}, the BspDP prior yields better results than the BDP in terms of median IAE in this example.  Consequently, our P-spline algorithm outperforms the BDP method as well.

%REMARK: PLEASE INSERT GRAPH WITH AR(1) AND AR(4) PSD ESTIMATES
\def\arraystretch{1.1}
% IAE
\begin{table}
	\centering
%	\begin{ruledtabular}
		\begin{tabular}{lccc}
			\toprule
			& $n=128$ & $n=256$ & $n=512$ \\ \hline
			AR(1)     &  &  &  \\ 
			B-spline &  0.870 & 0.714 & 0.594  \\
			\textit{Equidistant knots}& & &   \\
			\hspace{0.5em}P-spline $d=1$ &  0.769 & 0.655 & 0.527 \\
			\hspace{0.5em}P-spline $d=2$&  0.698 & 0.609 & 0.629 \\
			\textit{Q-spaced knots}& & &   \\
			\hspace{0.5em}P-spline $d=1$ & 0.848 & 0.771 & 0.614 \\
			\hspace{0.5em}P-spline $d=2$ & 0.920 & 0.840 & 0.648 \\ \hline
			AR(4)     &  &  &  \\ 
			B-spline & 2.990 & 2.202 & 1.800   \\
			\textit{Equidistant knots}& & &   \\
			\hspace{0.5em}P-spline $d=1$ &  2.905 & 2.389 & 2.306\\
			\hspace{0.5em}P-spline $d=2$&  3.149 & 2.566 & 2.387\\
			\textit{Q-spaced knots}& & &   \\
			\hspace{0.5em}P-spline $d=1$ & 2.517 & 1.843 & 1.486\\
			\hspace{0.5em}P-spline $d=2$ &  2.750 & 1.989 & 1.673\\
			\bottomrule
		\end{tabular}
%	\end{ruledtabular}
	\caption{Median IAE for the psd estimates using the BspDP and the P-spline priors for simulated AR(1) and AR(4) time series with  different penalty orders ($d$) and knot location schemes for the P-splines.}
	\label{table:sim_IAE}
\end{table}

%%%%%%%%%%%%%%%%%%%%%%%%%%%%
%%% Coverage probability %%%
%%%%%%%%%%%%%%%%%%%%%%%%%%%%

The IAE measures the accuracy of the  point estimates of the spectral density. In the following, we will also compare the coverage of posterior uniform credible bands.
%but we could be interested in testing interval estimates.  To assess this estimation, the uniform credible band is considered.  This approach allows to measure the coverage levels for the %whole spectral density. 
We define the $100(1-\alpha)\%$ uniform credible band as
\begin{align*}
\widehat{f}(\lambda) \pm \zeta_\alpha \times \text{mad} \left( \widehat{f}_i (\lambda) \right), \:\: \lambda \in [0,\pi],
\end{align*}
where $\widehat{f}(\lambda)$ is the pointwise posterior median spectral density, $\text{mad}( \widehat{f}_i (\lambda))$ is the median absolute deviation of the posterior samples $\widehat{f}_i (\lambda)$ and $\zeta_\alpha$ is such that
\begin{align*}
\text{P}\left( \max \left\{ \dfrac{|\widehat{f}_i (\lambda) - \widehat{f} (\lambda)|}{\text{mad}(\widehat{f}_i (\lambda))} \right\} \leq \zeta_{\alpha} \right) = 1 - \alpha.
\end{align*}

The idea is to measure the proportion of times that the true spectral density is fully contained by the uniform credible band in the 300 replications.  The results are displayed in Table~\ref{table:sim_covPbb}.  In addition, we calculate the proportion of times that the uniform credible band contains the value of the true psd
at each of the Fourier frequencies.  The median of these pointwise coverage proportions are displayed in Table~\ref{table:sim_covPbb_median}.

For the AR(1) model, the BspDP and first order penalty P-spline 90\% credible bands cover the whole true spectral density function, i.e.\ for all frequencies $\lambda$, in 
almost all the 300 independent analyses.  In general, the latter method shows diminishing  performance as the penalty degree increases.  For the AR(4) model, the P-spline method with equidistant knots performs poorly.  On the other hand, the quantile-based knot scheme yields evidently much better results, which are comparable (in the case of the first order penalty) to those obtained via the BspDP algorithm.  %For $n=\{128,256\}$, the latter has higher coverage probabilities than our proposal, except for $n=512$.  
For the P-spline priors, the coverage probabilities decrease as the penalty degree order increases, which can be explained by the forced smoothness imposed by the higher penalty order.  As shown by \cite{Edwards2019}, the BspDP algorithm produces better results than the BDP prior in the AR(4) case, which coverage probability is zero across $n=\{128,256,512\}$. Consequently, our P-spline algorithm with Q-spaced knots outperforms the BDP method in this case.

Even though the P-spline method with equidistant knots for the AR(4) case performs poorly in terms of uniform coverage, i.e. the proportion of times that the credible band covers  the true psd {\em completely} in the 300 runs, it is important to note that this is only because the uniform band does not cover
the psd for  just very few frequencies. As is evident from 
Table~\ref{table:sim_covPbb_median},  only for a small fraction of the frequencies the psd is not covered by the uniform credible band.  In this context, BspDP and Q-spaced  knot P-spline algorithms perform equally well.  

%For the AR(4) model, the P-spline prior with P-spaced knots has the best coverage across the different specifications. For the equidistant knots the coverage is always greater than 90\%.  For the first penalty order P-splines, the coverage probabilities are all higher than the ones produced by the BspDP prior algorithm.  For the second order penalty, it has a better performance, except for $n=512$, where the values are equal.  For the third order penalty, we see a mix of results, that is equal, higher and lower values for the increasing sequence of $n$, respectively.  For the P-spline priors, in particular for $n=512$, the coverage probabilities decrease as the penalty degree order increases, which can be explained by the forced smoothness imposed by the higher penalty order.  In general, P-spline priors have a better coverage probability than the BspDP method which as shown by  \cite{Edwards2019} have better coverage probabilities than the BDP prior in the AR(4) case.

% Coverage probability Table
\begin{table}
	\centering
%	\begin{ruledtabular}
		\begin{tabular}{lccc}
			\toprule
			& $n=128$ & $n=256$ & $n=512$ \\ \hline
			AR(1)     &  &  &  \\ 
			B-spline &  1.000 & 1.000 & 0.997 \\
			\textit{Equidistant}& & &   \\
			\hspace{0.5em}P-spline $d=1$&  1.000 & 1.000 & 0.997\\
			\hspace{0.5em}P-spline $d=2$& 0.993 & 1.000 & 0.993  \\
			\textit{Q-spaced knots}& & &   \\
			\hspace{0.5em}P-spline $d=1$ & 1.000 & 1.000 & 0.997 \\
			\hspace{0.5em}P-spline $d=2$ & 0.933 & 0.527 & 0.713 \\ \hline		       
			AR(4)     &  &  &  \\ 
			B-spline &  0.907 & 0.970 & 0.897  \\
			\textit{Equidistant knots}& & &   \\
			\hspace{0.5em}P-spline $d=1$ &  0.547 & 0.343 & 0.010 \\
			\hspace{0.5em}P-spline $d=2$ & 0.320 & 0.170 & 0.007\\
			\textit{Q-spaced knots}& & &   \\	
			\hspace{0.5em}P-spline $d=1$ & 0.800 & 0.833 & 0.803 \\
			\hspace{0.5em}P-spline $d=2$ & 0.313 & 0.363 & 0.277 \\
			\bottomrule
		\end{tabular}
%	\end{ruledtabular}
	\caption{Proportion of times that the true psd is entirely encapsulated within the 90\% uniform credible band of the 300 estimates.}
	\label{table:sim_covPbb}
\end{table}

\begin{table}
	\centering
		\begin{tabular}{lccc}
			\toprule
			& $n=128$ & $n=256$ & $n=512$ \\ \hline
			AR(1)     &  &  &  \\ 
			B-spline &  1.000 & 1.000 & 1.000 \\
			\textit{Equidistant \&}& & &   \\
			\textit{Q-spaced knots}& & &   \\
			\hspace{0.5em}P-spline $d=1,2$ & 1.000 & 1.000 & 1.000  \\ \hline	 		      
			AR(4)     &  &  &  \\ 
			B-spline &  1.000 & 1.000 & 1.000 \\
			\textit{Equidistant knots}& & &   \\
			\hspace{0.5em}P-spline $d=1$ &  1.000 & 0.992 & 0.984 \\
			\hspace{0.5em}P-spline $d=2$&  0.984 & 0.984 & 0.980 \\
			\textit{Q-spaced knots}& & &   \\	
			\hspace{0.5em}P-spline $d=1$ & 1.000 & 1.000 & 1.000 \\
			\hspace{0.5em}P-spline $d=2$ & 0.984 & 0.992 & 0.992\\	
			\bottomrule
		\end{tabular}
	\caption{Median of the pointwise  proportions that the true psd at each Fourier frequency is contained within the $90\%$ uniform credible bands.}
	\label{table:sim_covPbb_median}
\end{table}

%%%%%%%%%%%%
%%% TIME %%%
%%%%%%%%%%%%

As \cite{Edwards2019} noted, one of the drawbacks of the BspDP prior  is its computational complexity relative to the BDP.  This is directly reflected in the run-time, which is approximately 2-3 times higher in this example (see Table~3 in \cite{Edwards2019}).  The median run-time in our analyses for the BspDP prior and the P-spline with Q-spaced knots are displayed in Table~\ref{table:sim_time}.  Note that the P-spline method based on equidistant knots has similar run-times to the Q-spaced knots.  The P-spline method is approximately 4-6 times faster than BspDP approach in these examples.  This is due to a reduced number of calculations  per iteration because the B-spline densities are calculated only  once at the start of the algorithm and there are no Dirichlet process approximations via the stick-breaking method.  

% TIME Table
\begin{table}
	\centering
%	\begin{ruledtabular}
		\begin{tabular}{lccc}
			\toprule
			& $n=128$ & $n=256$ & $n=512$ \\ \hline
			AR(1)     &  &  &  \\ 
			B-spline &   24.323 & 28.258 & 34.093 \\
			\textit{Q-spaced knots}& & &   \\
			\hspace{0.5em}P-spline $d=1$&  4.859 & 7.450 & 9.606\\
			\hspace{0.5em}P-spline $d=2$&  4.538 & 6.863 & 8.783\\
			B-spline/P-spline &  6.090 & 4.368 & 4.493 \\ \hline
			AR(4)     &  &  & \\ 
			B-spline & 24.983 & 29.796 & 36.913  \\
			\textit{Q-spaced knots}& & &   \\
			\hspace{0.5em}P-spline $d=1$ & 4.883 & 7.453 & 9.665 \\
			\hspace{0.5em}P-spline $d=2$& 4.543 & 6.887 & 8.827 \\
			B-spline/P-spline & 6.062 & 4.591 & 5.523 \\	
			\bottomrule
		\end{tabular}
%	\end{ruledtabular}
	\caption{Median run-times in minutes and relative run-times with respect to an average of the P-spline run-times.}
	\label{table:sim_time}
\end{table}

In general terms, the P-spline algorithm with our novel quantile-based  knot allocation scheme offers the best trade-off between accuracy and computation cost.  The  rule-of-thumb criterion,  $K=\min\left\{n/4, 40\right\}$, for the number of B-spline densities in the P-spline prior algorithm worked perfectly well in this example. 

%Overall, the P-spline prior outperforms the B-spline prior in this example in terms of IAE, coverage probability and run-time.  Particularly, our novel knot allocation scheme, the periodogram-spaced knots, yield the best results in the case of peaked spectral densities.  The  rule-of-thumb criterion,  $K=\min\left\{n/4, 40\right\}$, for the number of 
%\textcolor{red}{P-splines} B-spline densities in the P-spline prior algorithm
% worked perfectly well in this example. 

\subsection{Sunspot data analysis}
Sunspots are regions of reduced surface temperature that are visible as darker spots on the sun's photosphere.
We analyse the average annual mean sunspot numbers for the year 1700-1987 consisting of  288 observations.  This is a classic dataset used to assess spectral density estimation methods.  A square root transformation is applied in order to make the observations more symmetrical and stationary. 

As in \cite{Edwards2019}, the MCMC algorithm based on the BspDP prior  is run for 100,000 iterations, with a burn-in period of 50,000 and thinning factor of 10.  Thus, 5,000 samples are used for the estimation.  This took about 22 minutes.  The other specifications are the same as those used in the simulation study.

The MCMC algorithm based on the P-spline prior is run in two stages.  In  the first stage, the proposal  is calibrated for the final run.  It consists of 25,000 iterations with a burn-in period of 10,000 and thinning factor of 10, resulting in 1,500 samples.  Then the algorithm is run for 75,000 iterations with a burn-in period of 25,000 and thinning factor of 10, resulting in 5,000 samples.  The analysis is performed using the first order penalty for equidistant and Q-spaced knots.   All other specifications are the same as those used in the simulation study.  A single analysis takes about 4.5 minutes.

\begin{figure}[]
	\centering
	\includegraphics[scale=0.40,clip=true,angle=0]{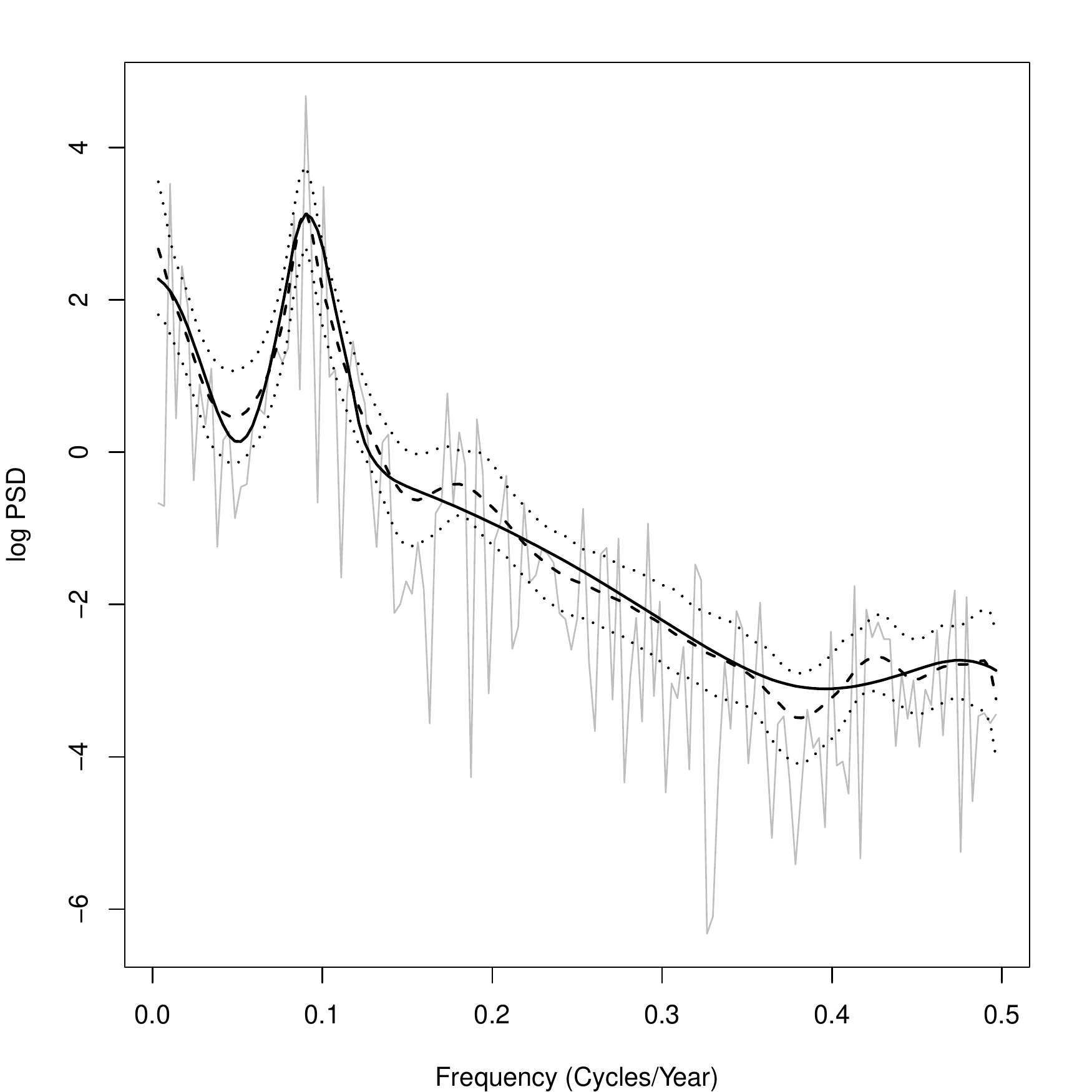}
	\caption{Log-spectral density estimate for the sunspot data. The continuous grey line represents the periodogram.  The continuous and dashed lines stand for the posterior median log psd obtained using the B-spline and P-spline priors, respectively.  The dot lines represents the 90\% pointwise credible bands for the latter prior, which consider a first order penalty and Q-spaced knots.}
	\label{fig:sunspot}
\end{figure}

%REMARK: PLEASE CHANGE THE TICK MARKS ON X-AXIS IF THESE ARE CYCLES/YEAR NOT ANGULAR FREQUENCIES.\smallskip

The results for the BspDP and P-spline prior  with first order penalty and Q-spaced knots are displayed in Figure~\ref{fig:sunspot} (p-spline with equidistant knot approach results are not shown due to their similarity to the Q-spaced knot estimates).  Both methods reveal a large peak at about 0.09. In other words, there is a periodic solar cycle of length $1/ 0.09 \approx 11$ years.  These results are consistent with those obtained via the BDP  prior~\citep{Choudhuri:2004}.
However, the BspDP estimation takes approximately 22 minutes compared to only 4.5 minutes for the the P-spline estimation. Thus the P-spline estimation is approximately
five times more efficient.

%REMARK: I WOULD  DELETE THE NEXT PARAGRAPH AND TABLE AS THIS SUGGESTS THAT THE BSPDP ESTIMATE IS THE GOLD STANDARD. BUT THE AR(4) SIMULATION STUDY HAS SHOWN THAT THIS IS NOT THE CASE. 
%We also estimate the 90\% credible bands obtained via the P-spline prior algorithm and calculate the proportion of times that these encapsulate the posterior median log psd obtained using the B-spline prior.  The results are displayed in Table~\ref{table:sunsont_knots}.  The coverage is quite high in general and decreases as the penalty order increases.  
%\begin{table}
%	\centering
%	\begin{ruledtabular}
%		\begin{tabular}{lccc}
%			\toprule
%			                   & $d=1$ & $d=2$ & $d=3$ \\ \hline
%			Equidistant knots  & 0.98 & 0.98 & 0.94 \\ 
%			Distributed knots  & 1.00 & 0.94 & 0.85  \\			
%			\bottomrule
%		\end{tabular}
%	\end{ruledtabular}
%	\caption{Proportion of times that the posterior median log psd obtained via B-spline prior algorithm is encapsulated within the 90\% uniform credible band obtained via the P-spline prior algorithm.}
%	\label{table:sunsont_knots}
%\end{table}

\subsection{Variable star S. Carinae data analysis}

\begin{figure}[]
	\centering
	\includegraphics[scale=0.4,clip=true,angle=0]{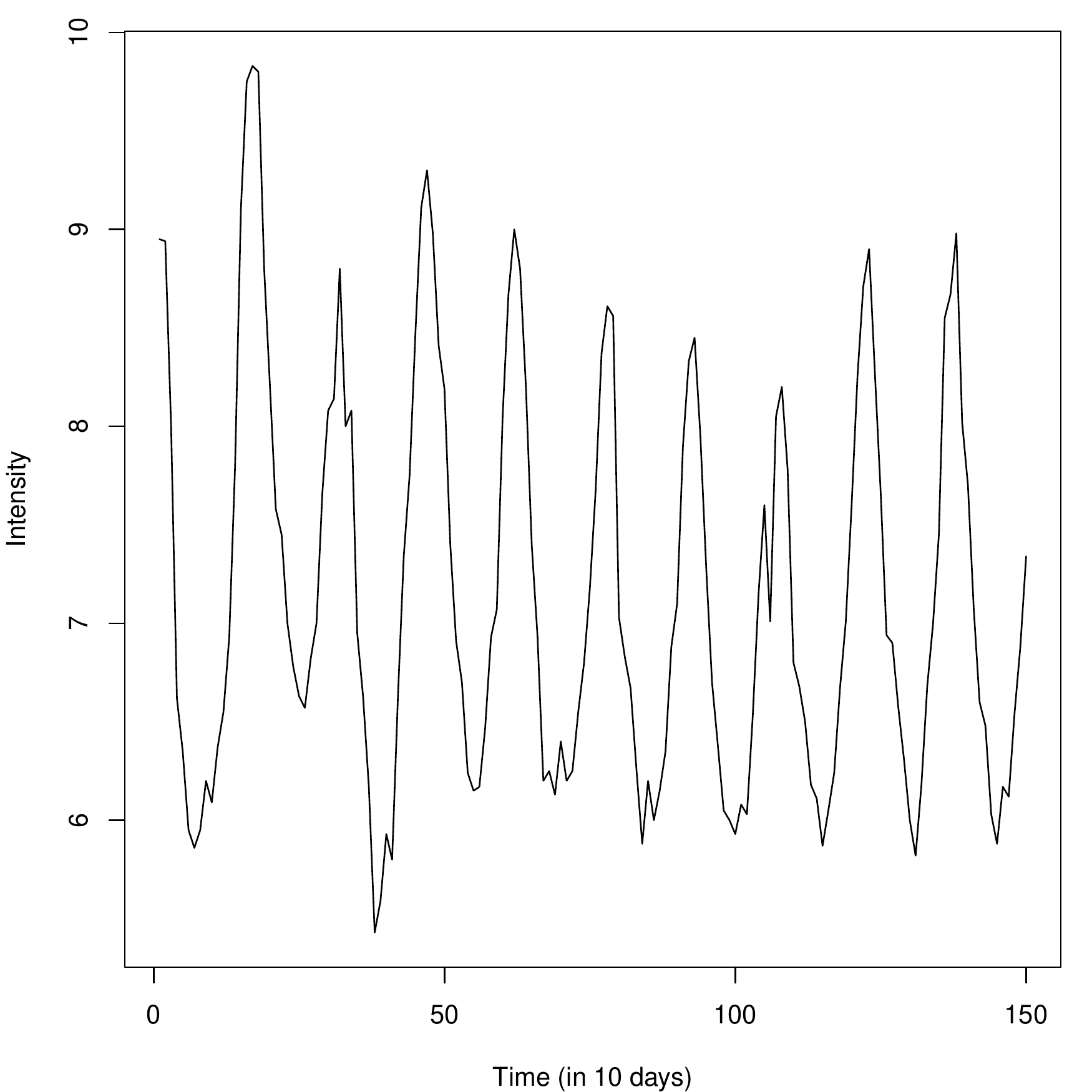}
	\caption{First 150 observations of the \textit{S. Carinae} data.}
	\label{fig:carinae_data}
\end{figure}

This dataset contains 1,189 visual observations of the {\it S.\ Carinae}, a variable star in the southern hemisphere sky.  These are daily observations collected by the Royal Astronomical Society of New Zealand, which correspond to 10 day averages of light intensities over several years.  This data set has been analysed in the literature previously (see for instance \cite{Cart:1997,Huerta:1999,Kirch:2018}).  It contains 40 missing observations that in this work have been replaced by the mean, thus not affecting the general features of the data.  In addition, the data has been squared root transformed and mean centred.  The first 150 observations are displayed in Figure~\ref{fig:carinae_data}, where the missing values are replaced by the mean.  

\cite{Kirch:2018} provided a psd estimate via a nonparametrically corrected (NPC) approach, which is able to detect several peaks in the function.  These features make this data set suitable to assess the impact of the knot location on the P-spline estimates. 

\begin{figure}[]
	\centering
	\includegraphics[scale=0.4,clip=true,angle=0]{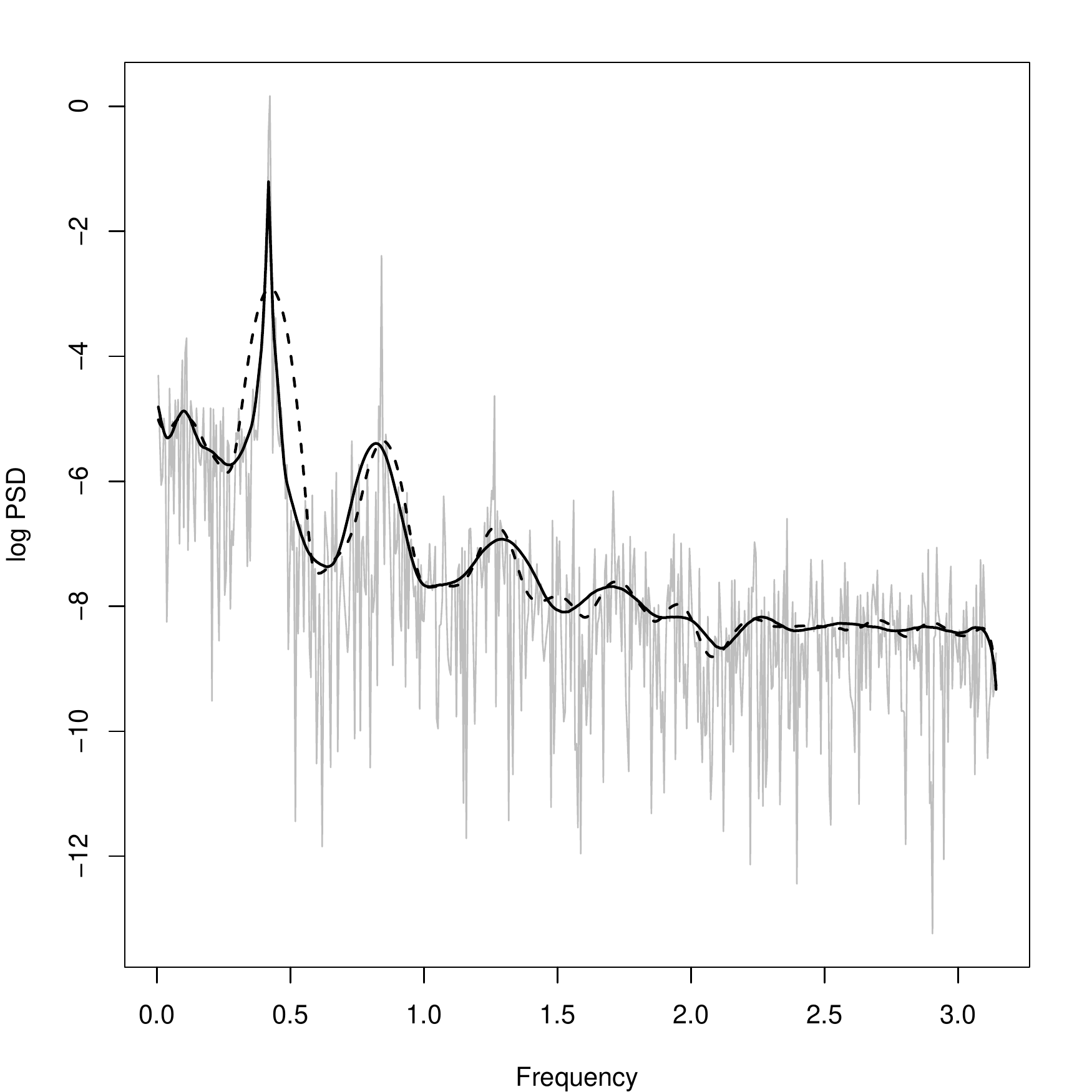}
	\caption{Log-spectral density estimate for the variable star {\it S.\ Carinae} data based on the P-spline prior. The continuous grey line represents the periodogram whereas the continuous and dashed black lines stand for the posterior median log psd obtained using Q-spaced and equidistant knots, respectively.}
	\label{fig:carinae}
\end{figure}

Again, the P-spline algorithm is run in two stages.  First, a Markov chain of 5,000 samples with a burn-in period of 1,000 and thinning factor of 10, resulting in 400 points, is used to calibrate the proposals of the final run.  Then, a Markov chain of 10,000 samples with a burn-in period of 2,000 and thinning factor of 10, that is 800 points, is used for the psd estimation.  We use 40 B-spline densities,  based on the rule-of-thumb criterion discussed above, and the first order penalty.  The whole process takes less than 1.5 minutes to run and is replicated for equidistant and Q-spaced knots. The results are displayed in Figure~\ref{fig:carinae}.

The periodogram shows several peaks with the main one located at around 0.42.  When the knots are equally spaced, the algorithm is unable to capture the sharpness of this peak,
in contrast to the Q-spaced knots.  This is because the periodogram-based knot allocation puts more knots in those regions which contain significant peaks as indicated by the periodogram.  The rest of peaks, which are smaller in comparison to the main one, are slightly better captured by the equidistant knots, since a large proportion of the Q-spaced knots have been allocated around the main peak. However, this can easily  be corrected by increasing the number of knots.  The psd estimate based on Q-spaced knots is comparable to the one obtained via the NPC approach \citep{Kirch:2018}.

%REMARK: TO SUPPORT THIS CLAIM, SHOULD WE INCLUDE ANOTHER GRAPH THAT COMPARES THE NPC ESTIMATE WITH P-SPACE P-SPLINES WITH LARGER NUMBER OF KNOTS?

\section{Discussion}

As shown by \cite{Edwards2019}, the BspDP prior outperforms the BDP prior in terms of IAE and uniform coverage probabilities for spiky psds.  This is explained by the local support of the B-splines and its better approximation properties \citep{Edwards2019}.  However, its performance is only achieved at a much higher computational cost.  These characteristics have motivated our P-spline approach.  

Although the P-spline prior is very similar to the B-spline prior as both are based on mixtures of B-spline distributions, there is quite a significant difference.  The latter works with a variable number and location of knots, which are driven by the data.  The P-spline prior assumes a fixed number of knots  with fixed locations, thus almost looks like a special case. However, the multivariate prior based on the difference penalty of the basis coefficients controls the smoothness of the spectral density, a feature that cannot be achieved with a univariate Dirichlet process prior. Most importantly, fixing the number and location of knots reduces the computational effort significantly, without sacrificing accuracy as shown in the application section.  To deal with abrupt peaks in the psd, we propose to locate the knots according to the peaks detected via the periodogram.  This approach does not affect the computational time and improves the estimates dramatically. %This also indicates that the BspDP computation time could be reduced by the choice of \textcolor{red}{Q-spaced} knots as initial values.

In our simulation study, we showed that the equidistant knot scheme for the P-spline method works quite well in the case of simple spectral structures (AR(1) model).  However, its performance is surpassed by the quantile-based  knot location scheme for complex structures (AR(4) model).  In this scenario particularly, we can highlight the good performance of the quantile-based knot scheme. This method outperforms the BspDP prior in terms of IAE and run-times, particularly for the complex psds.    Regarding uniform coverage,  a {\em first order} penalty P-spline coupled with
Q-spaced knots works best. For this combination, the uniform coverage almost reaches the nominal level.

%In our simulation study, we showed that, in general, the P-spline prior with \textcolor{red}{Q-spaced} knots outperforms the BspDP prior in terms of IAE and run-times, particularly for the complex psd.  However, the results were not conclusive for coverage probabilities.  From this study, it seems that a {\em first order} penalty P-spline works reasonably well.

We also assessed our proposal in the classic sunspot dataset, in which the posterior distribution based on the BspDP prior %B-spline prior algorithm
 is able to estimate correctly the solar cycle that occurs every 11.07 years.  The posterior distribution based on the P-spline prior yields almost identical results, but in  significantly  less computational time.  

Finally, we assess the fit of the posterior distribution based on the P-spline prior a under two different knot location schemes: equally spaced and Q-spaced knots.  For this, we estimated the psd for the {\it S.\ Carinae} time series which has a spectral density with several sharp peaks.  Equidistant knots failed to capture the main peak of this spectral density whereas Q-spaced knots significantly improved  the results without affecting the computational time.

The number of B-spline densities was selected according to the criterion proposed by \cite{Ruppert2002}.  It seems to work well for simple spectral densities.  However, more complex functions could require larger number of B-spline densities.  This can be evaluated studying the peaks of the periodogram.  Our knot location proposal is very useful in these situations.  Future research will explore whether the choice of the number of knots can be related to the entropy of the periodogram.

As already reviewed in the introduction, many Bayesian nonparametric approaches to spectral density estimation propose to put a prior on the log-spectral density. This could be easily implemented by modifying (\ref{eq:ps_prior}) to
 \[ \log f(\pi \omega)=\sum_{k=1}^K v_k b_{k,r}(\omega,{\bm \xi})\]
and it could be interesting to examine for what class of spectral densities this prior specification would yield better posterior inference. 

Another interesting question is whether the P-spline prior can be extended to enable a Bayesian analysis of multivariate time series. In this case, the spectral density is a 
Hermitian positive definite  matrix-valued function. A novel Bayesain nonparametric approach to multivariate spectral density estimation was proposed by
\cite{Meier2020} based on a generalization of the Bernstein-Dirchlet process prior where Bernstein polynomials are used to smooth a matrix-Gamma process.
An avenue for future research will be to explore whether a computational speed-up might be possible by making use of P-splines in smoothing the elements of the
Cholesky decomposition of the spectral density matrix.

In conclusion, the P-spline prior with the quantile-based knot placement scheme offers a more computationally viable alternative to the BspDP prior for spectral density estimation.  As shown in this work, it can handle spectral densities with several peaks in a reasonable run-time.

The \textsf{R} package \texttt{psplinePsd} can be downloaded from \href{https://github.com/pmat747/psplinePsd}{GitHub}.
%The \textsf{R} package \texttt{psplinePsd} will be available on CRAN in due course.  In the meanwhile, it can be downloaded from \href{https://github.com/pmat747/psplinePsd}{GitHub}. 

\begin{acknowledgements}
We thank Thomas Yee for helpful discussions on P-splines.	
We also thank the New Zealand eScience Infrastructure (NeSI) for their high performance computing facilities, and the Centre for eResearch at the University of Auckland for their technical support. PM's and RM's work is supported by Grant 3714568 from the University of Auckland Faculty Research Development Fund and the DFG Grant KI 1443/3-1. RM gratefully acknowledges support by a James Cook Fellowship from Government funding, administered by the Royal Society Te Ap\={a}rangi.  All analysis was conducted in \textsf{R}, an open-source statistical software available on \textsf{CRAN} (cran.r-project.org).
\end{acknowledgements}

\bibliographystyle{plainnat}
\bibliography{pSplinesReferences}  % Name of BibTeX file

\begin{thebibliography}{38}
\providecommand{\natexlab}[1]{#1}
\providecommand{\url}[1]{\texttt{#1}}
\expandafter\ifx\csname urlstyle\endcsname\relax
  \providecommand{\doi}[1]{doi: #1}\else
  \providecommand{\doi}{doi: \begingroup \urlstyle{rm}\Url}\fi

\bibitem[Bremhorst and Lambert(2016)]{Bremhorst:2016}
Vincent Bremhorst and Philippe Lambert.
\newblock Flexible estimation in cure survival models using {B}ayesian
  p-splines.
\newblock \emph{Computational Statistics \& Data Analysis}, 93:\penalty0 270 --
  284, 2016.
\newblock ISSN 0167-9473.
\newblock \doi{https://doi.org/10.1016/j.csda.2014.05.009}.
\newblock URL
  \url{http://www.sciencedirect.com/science/article/pii/S0167947314001492}.

\bibitem[Brockwell and Davis(1991)]{Brockwell:1986}
Peter~J Brockwell and Richard~A Davis.
\newblock \emph{Time Series: Theory and Methods}.
\newblock Springer-Verlag New York, New York, USA, 2 edition, 1991.
\newblock ISBN 978-0-387-97429-3.
\newblock \doi{10.1007/978-1-4419-0320-4}.
\newblock URL \url{https://doi.org/10.1007/978-1-4419-0320-4}.

\bibitem[Cadonna et~al.(2017)Cadonna, Kottas, and Prado]{Cadonna2017}
A.~Cadonna, A.~Kottas, and R.~Prado.
\newblock Bayesian mixture modeling for spectral density estimation.
\newblock \emph{Statistics \& Probability Letters}, 125:\penalty0 189--195,
  2017.
\newblock ISSN 0167-7152.
\newblock \doi{https://doi.org/10.1016/j.spl.2017.02.008}.
\newblock URL
  \url{http://www.sciencedirect.com/science/article/pii/S0167715217300573}.

\bibitem[Carter and Kohn(1997)]{Cart:1997}
C.~K. Carter and R.~Kohn.
\newblock Semiparametric {B}ayesian inference for time series with mixed
  spectra.
\newblock \emph{Journal of the Royal Statistical Society, Series B},
  59\penalty0 (1):\penalty0 255--268, 1997.
\newblock \doi{10.1111/1467-9868.00067}.
\newblock URL \url{https://doi.org/10.1111/1467-9868.00067}.

\bibitem[Choudhuri et~al.(2004)Choudhuri, Ghosal, and Roy]{Choudhuri:2004}
Nidhan Choudhuri, Subhashis Ghosal, and Anindya Roy.
\newblock Bayesian estimation of the spectral density of a time series.
\newblock \emph{Journal of the American Statistical Association}, 99\penalty0
  (468):\penalty0 1050--1059, 2004.
\newblock \doi{10.1198/016214504000000557}.
\newblock URL \url{https://doi.org/10.1198/016214504000000557}.

\bibitem[Edwards et~al.(2019)Edwards, Meyer, and Christensen]{Edwards2019}
Matthew Edwards, Renate Meyer, and Nelson Christensen.
\newblock Bayesian nonparametric spectral density estimation using b-spline
  priors.
\newblock \emph{Statistics and Computing}, 29\penalty0 (1):\penalty0 67--78,
  2019.
\newblock ISSN 0960-3174.
\newblock URL \url{https://doi.org/10.1007/s11222-017-9796-9}.

\bibitem[Edwards et~al.(2018)Edwards, Meyer, and
  Christensen]{Edwards:bsplinePsd:2018}
Matthew~C. Edwards, Renate Meyer, and Nelson Christensen.
\newblock \emph{bsplinePsd: Bayesian nonparametric spectral density estimation
  using b-spline priors}, 2018.
\newblock URL \url{https://CRAN.R-project.org/package=bsplinePsd}.
\newblock {R package version 0.6.0}.

\bibitem[Eilers and Marx(1996)]{Eilers:1996}
Paul H.~C. Eilers and Brian~D. Marx.
\newblock Flexible smoothing with b -splines and penalties.
\newblock \emph{Statist. Sci.}, 11\penalty0 (2):\penalty0 89--121, 05 1996.
\newblock \doi{10.1214/ss/1038425655}.
\newblock URL \url{https://doi.org/10.1214/ss/1038425655}.

\bibitem[Eilers et~al.(2015)Eilers, Marx, and Durb{\'a}n]{Eilers2015}
Paul H.~C. Eilers, Brian~D. Marx, and Mar{\'i}a Durb{\'a}n.
\newblock Twenty years of p-splines.
\newblock \emph{SORT: statistics and operations research transactions},
  39\penalty0 (2), 2015.
\newblock ISSN 1696-2281.
\newblock URL \url{http://hdl.handle.net/2117/88526}.

\bibitem[Gangopadhyay et~al.(1999)Gangopadhyay, Mallick, and
  Denison]{Gangopadhyay:1999}
A.K. Gangopadhyay, B.K. Mallick, and D.G.T. Denison.
\newblock Estimation of spectral density of a stationary time series via an
  asymptotic representation of the periodogram.
\newblock \emph{Journal of Statistical Planning and Inference}, 75\penalty0
  (2):\penalty0 281 -- 290, 1999.
\newblock ISSN 0378-3758.
\newblock \doi{https://doi.org/10.1016/S0378-3758(98)00148-7}.
\newblock URL
  \url{http://www.sciencedirect.com/science/article/pii/S0378375898001487}.

\bibitem[Green(1995)]{Green:1995}
Peter~J. Green.
\newblock {Reversible jump Markov chain Monte Carlo computation and Bayesian
  model determination}.
\newblock \emph{Biometrika}, 82\penalty0 (4):\penalty0 711--732, 12 1995.
\newblock ISSN 0006-3444.
\newblock \doi{10.1093/biomet/82.4.711}.
\newblock URL \url{https://dx.doi.org/10.1093/biomet/82.4.711}.

\bibitem[Huerta and West(1999)]{Huerta:1999}
G.~Huerta and M.~West.
\newblock Bayesian inference on periodicities and component spectral structure
  in time series.
\newblock \emph{Journal of Time Series Analysis}, 20\penalty0 (4):\penalty0
  401--416, 1999.
\newblock \doi{10.1111/1467-9892.00145}.
\newblock URL \url{https://doi.org/10.1111/1467-9892.00145}.

\bibitem[Jullion and Lambert(2007)]{Jullion:2007}
Astrid Jullion and Philippe Lambert.
\newblock Robust specification of the roughness penalty prior distribution in
  spatially adaptive {B}ayesian p-splines models.
\newblock \emph{Computational Statistics \& Data Analysis}, 51\penalty0
  (5):\penalty0 2542 -- 2558, 2007.
\newblock ISSN 0167-9473.
\newblock \doi{https://doi.org/10.1016/j.csda.2006.09.027}.
\newblock URL
  \url{http://www.sciencedirect.com/science/article/pii/S0167947306003549}.

\bibitem[Kauermann and Opsomer(2011)]{Kauermann2011}
G{\"o}ran Kauermann and Jean~D. Opsomer.
\newblock Data-driven selection of the spline dimension in penalized spline
  regression.
\newblock \emph{Biometrika}, 98\penalty0 (1):\penalty0 225--230, 2011.
\newblock ISSN 00063444.
\newblock URL \url{http://www.jstor.org/stable/29777177}.

\bibitem[Kirch et~al.(2018)Kirch, Edwards, Meier, and Meyer]{Kirch:2018}
Claudia Kirch, Matthew~C. Edwards, Alexander Meier, and Renate Meyer.
\newblock Beyond {W}hittle: Nonparametric correction of a parametric likelihood
  with a focus on {B}ayesian time series analysis.
\newblock \emph{Bayesian Anal.}, 2018.
\newblock \doi{10.1214/18-BA1126}.
\newblock URL \url{https://doi.org/10.1214/18-BA1126}.
\newblock Advance publication.

\bibitem[Krivobokova et~al.(2006)Krivobokova, Kauermann, and
  Archontakis]{Krivobokova}
Tatyana Krivobokova, Göran Kauermann, and Theofanis Archontakis.
\newblock Estimating the term structure of interest rates using penalized
  splines.
\newblock \emph{Statistical Papers}, 47\penalty0 (3):\penalty0 443--459, 2006.
\newblock ISSN 0932-5026.

\bibitem[Lambert(2007)]{Lambert:2007}
Philippe Lambert.
\newblock Archimedean copula estimation using {B}ayesian splines smoothing
  techniques.
\newblock \emph{Computational Statistics \& Data Analysis}, 51\penalty0
  (12):\penalty0 6307 -- 6320, 2007.
\newblock ISSN 0167-9473.
\newblock \doi{https://doi.org/10.1016/j.csda.2007.01.018}.
\newblock URL
  \url{http://www.sciencedirect.com/science/article/pii/S0167947307000217}.

\bibitem[Lang and Brezger(2004)]{Lang:2004}
Stefan Lang and Andreas Brezger.
\newblock Bayesian p-splines.
\newblock \emph{Journal of Computational and Graphical Statistics}, 13\penalty0
  (1):\penalty0 183--212, 2004.
\newblock \doi{10.1198/1061860043010}.
\newblock URL \url{https://doi.org/10.1198/1061860043010}.

\bibitem[Likhachev(2017)]{Likhachev2017}
D.V. Likhachev.
\newblock Selecting the right number of knots for b-spline parameterization of
  the dielectric functions in spectroscopic ellipsometry data analysis.
\newblock \emph{Thin Solid Films}, 636:\penalty0 519 -- 526, 2017.
\newblock ISSN 0040-6090.
\newblock \doi{https://doi.org/10.1016/j.tsf.2017.06.056}.
\newblock URL
  \url{http://www.sciencedirect.com/science/article/pii/S0040609017304911}.

\bibitem[Maturana-Russel and Meyer(2020)]{psplinepackage}
Patricio Maturana-Russel and Renate Meyer.
\newblock \emph{psplinePsd: P-splines for spectral density estimation}, 2020.
\newblock URL \url{https://github.com/pmat747/psplinePsd}.

\bibitem[Meier et~al.(2020)Meier, Kirch, and Meyer]{Meier2020}
Alexander Meier, Claudia Kirch, and Renate Meyer.
\newblock Bayesian nonparametric analysis of multivariate time series: A matrix
  gamma process approach.
\newblock \emph{Journal of Multivariate Analysis}, 175, 2020.
\newblock ISSN 0047-259X.

\bibitem[Pawitan and O'sullivan(1994)]{Pawitan}
Yudi Pawitan and Finbarr O'sullivan.
\newblock Nonparametric spectral density estimation using penalized whittle
  likelihood.
\newblock \emph{Journal of the American Statistical Association}, 89\penalty0
  (426):\penalty0 600--610, 1994.
\newblock \doi{10.1080/01621459.1994.10476785}.
\newblock URL \url{https://doi.org/10.1080/01621459.1994.104767}.

\bibitem[Pensky et~al.(2007)Pensky, Vidakovic, , and De~Canditiis]{Pensky:2007}
Marianna Pensky, Brani Vidakovic, , and Daniela De~Canditiis.
\newblock Bayesian decision theoretic scale-adaptive estimation of a
  log-spectral density.
\newblock \emph{Statistica Sinica}, 17:\penalty0 635--666, 2007.
\newblock URL
  \url{http://www3.stat.sinica.edu.tw/statistica/j17n2/j17n210/j17n210.html}.

\bibitem[Perron and Mengersen(2001)]{Perron:2001}
F.~Perron and K.~Mengersen.
\newblock Bayesian nonparametric modeling using mixtures of triangular
  distributions.
\newblock \emph{Biometrics}, 57\penalty0 (2):\penalty0 518--528, 2001.
\newblock ISSN 0006341X, 15410420.
\newblock URL \url{http://www.jstor.org/stable/3068361}.

\bibitem[Petrone(1999{\natexlab{a}})]{Petrone:1999a}
Sonia Petrone.
\newblock Random {B}ernstein polynomials.
\newblock \emph{Scandinavian Journal of Statistics}, 26\penalty0 (3):\penalty0
  373--393, 1999{\natexlab{a}}.
\newblock \doi{10.1111/1467-9469.00155}.
\newblock URL
  \url{https://onlinelibrary.wiley.com/doi/abs/10.1111/1467-9469.00155}.

\bibitem[Petrone(1999{\natexlab{b}})]{Petrone:199b}
Sonia Petrone.
\newblock Bayesian density estimation using {B}ernstein polynomials.
\newblock \emph{The Canadian Journal of Statistics / La Revue Canadienne de
  Statistique}, 27\penalty0 (1):\penalty0 105--126, 1999{\natexlab{b}}.
\newblock ISSN 03195724.
\newblock URL \url{http://www.jstor.org/stable/3315494}.

\bibitem[Polson et~al.(2013)Polson, Scott, and Windle]{Polson:2013}
Nicholas~G. Polson, James~G. Scott, and Jesse Windle.
\newblock Bayesian inference for logistic models using {P}\'olya–{G}amma
  latent variables.
\newblock \emph{Journal of the American Statistical Association}, 108\penalty0
  (504):\penalty0 1339--1349, 2013.
\newblock \doi{10.1080/01621459.2013.829001}.
\newblock URL \url{https://doi.org/10.1080/01621459.2013.829001}.

\bibitem[Ramsay et~al.(2020)Ramsay, Wickham, Graves, and Hooker]{fda}
J.~O. Ramsay, Hadley Wickham, Spencer Graves, and Giles Hooker.
\newblock \emph{fda: Functional Data Analysis}, 2020.
\newblock URL \url{https://CRAN.R-project.org/package=fda}.
\newblock R package version 2.4.8.1.

\bibitem[Rodríguez-Álvarez et~al.()Rodríguez-Álvarez, Durban, Lee, and
  Eilers]{Rodriguez}
María Rodríguez-Álvarez, Maria Durban, Dae-Jin Lee, and Paul Eilers.
\newblock On the estimation of variance parameters in non-standard generalised
  linear mixed models: application to penalised smoothing.
\newblock \emph{Statistics and Computing}, 29\penalty0 (3):\penalty0 483--500.
\newblock ISSN 0960-3174.

\bibitem[Rosen et~al.(2012)Rosen, Wood, and Stoffer]{Rosen:2012}
Ori Rosen, Sally Wood, and David~S. Stoffer.
\newblock Adaptspec: Adaptive spectral estimation for nonstationary time
  series.
\newblock \emph{Journal of the American Statistical Association}, 107\penalty0
  (500):\penalty0 1575--1589, 2012.
\newblock ISSN 0162-1459.
\newblock \doi{10.1080/01621459.2012.716340}.
\newblock URL \url{https://doi.org/10.1080/01621459.2012.716340}.

\bibitem[Ruppert(2002)]{Ruppert2002}
David Ruppert.
\newblock Selecting the number of knots for penalized splines.
\newblock \emph{Journal of Computational and Graphical Statistics}, 11\penalty0
  (4):\penalty0 735--757, 2002.
\newblock \doi{10.1198/106186002853}.
\newblock URL \url{https://doi.org/10.1198/106186002853}.

\bibitem[Sethuraman(1994)]{Sethuraman:1994}
Jayaram Sethuraman.
\newblock A constructive definition of {Dirichlet} priors.
\newblock \emph{Statistica Sinica}, 4:\penalty0 639--650, 1994.
\newblock URL
  \url{http://www3.stat.sinica.edu.tw/statistica/j4n2/j4n216/j4n216.htm}.

\bibitem[Shao and Wu(2007)]{ShaoXiaofeng2007ASTf}
Xiaofeng Shao and Wei~Biao Wu.
\newblock Asymptotic spectral theory for nonlinear time series.
\newblock \emph{The Annals of Statistics}, 35\penalty0 (4):\penalty0
  1773--1801, 2007.
\newblock ISSN 00905364.
\newblock URL \url{http://www.jstor.org/stable/25464559}.

\bibitem[Wand and Ormerod(2008)]{WandOrmerod}
M.~P. Wand and J.~T. Ormerod.
\newblock On semiparametric regression with {O'S}ullivan penalized splines.
\newblock \emph{Australian \& New Zealand Journal of Statistics}, 50\penalty0
  (2):\penalty0 179--198, 2008.
\newblock ISSN 1369-1473.

\bibitem[Wegener and Kauermann(2017)]{Wegener}
Michael Wegener and Goran Kauermann.
\newblock Forecasting in nonlinear univariate time series using penalized
  splines.(report).
\newblock \emph{Statistical Papers}, 58\penalty0 (3):\penalty0 557, 2017.
\newblock ISSN 0932-5026.

\bibitem[Whittle(1957)]{Whittle:1957}
P.~Whittle.
\newblock Curve and periodogram smoothing.
\newblock \emph{Journal of the Royal Statistical Society. Series B
  (Methodological)}, 19\penalty0 (1):\penalty0 38--63, 1957.
\newblock ISSN 00359246.
\newblock URL \url{http://www.jstor.org/stable/2983994}.

\bibitem[Wood(2017)]{WoodSimon2017Pwdb}
Simon Wood.
\newblock P-splines with derivative based penalties and tensor product
  smoothing of unevenly distributed data.
\newblock \emph{Statistics and Computing}, 27\penalty0 (4):\penalty0 985--989,
  2017.
\newblock ISSN 0960-3174.

\bibitem[Wood and Fasiolo(2017)]{Wood2017}
Simon~N. Wood and Matteo Fasiolo.
\newblock A generalized fellner‐schall method for smoothing parameter
  optimization with application to tweedie location, scale and shape models.
\newblock \emph{Biometrics}, 73\penalty0 (4):\penalty0 1071--1081, 2017.
\newblock ISSN 0006-341X.

\end{thebibliography}

\end{document}